\documentstyle[12pt,a4]{article}
\begin{document}
\title { ON A CLASS OF RATIONAL AND MIXED SOLITON-RATIONAL SOLUTIONS 
OF TODA LATTICE}
\author{A. S. C\^arstea\thanks{e-mail: acarst@theor1.ifa.ro}\and
D.Grecu\thanks{e-mail: dgrecu@roifa.ifa.ro}\and{\it Institute of Atomic Physics
Bucharest, MG 6, Romania}}

\maketitle

\begin{abstract}
A class of rational solutions of Toda lattice
satisfying certain Backlund transformations and 
a class of mixed rational-soliton solutions (quasisolitons)in wronskian form
are obtained using the method of Ablowitz and Satsuma. Also an extended class
of rational solutions are found using an appropriate recursion relation.
They are also solutions of Boussinesq equation and it is conjectured
that there is a larger class of common solutions of both equations.
\end{abstract}   

\section{Introduction}
The class of rational solutions was firstly investigated for KdV equation
\cite{air}, \cite{adl}. A very simple way to find them was
developed by Ablowitz and Satsuma (AS) \cite{as1}
 and consists in taking the "long wave limit" in the multisoliton solution.
The method was used successfully to obtain the rational solutions also for
other nonlinear completely integrable systems \cite{as1}, \cite{as2}.

Mixed rational-soliton solutions (quasisolitons) of KdV equation were
discovered by Ablowitz and Cornille \cite{ac}
 and later were studied by H.Airault and
M.J.Ablowitz \cite{aair}.
 They were also found using the limiting procedure of AS \cite{grci}.

In spite of the large popularity of Toda lattice (TL)
 as the most known and
studied  nonlinear integrable lattice model
\cite{toda1}
 these types of solutions were very little investigated. In a previous paper 
following AS procedure we obtained the expressions for the first rational
and mixed rational-soliton solutions \cite{carstea}.

Let us summarize the main results of \cite{carstea} :

\begin{itemize}
\item one rational solution:

\begin{equation}\label{zakrs1}
f_{n}^{(1)}=(n-\epsilon t)
\end{equation}
where $\epsilon = \pm 1$ depending on the propagation direction of the soliton.
\item two rational solution:

\begin{equation}\label{zakrs2-}
f_{n}^{(2)}=n^{2}-t^{2}-1/4
\end{equation}
		
if the two solitons are moving in opposite directions,
and
\begin{equation}\label{zakrs2+}
f_{n}^{(2)}=4(n-\epsilon t)^{3}-(n-\epsilon t)+2 \epsilon t
\end{equation}
if they move in the same direction.
\item mixed 1-rational 1-soliton solution:
\begin{equation}\label{zakqs}
f_{n}^{(1,1)}=(n-\epsilon t)+(n-\epsilon t+B)\exp(-2 \eta)
\end{equation}
where
\begin{equation}
\eta =Pn-\epsilon't \sinh{P} +\eta^{(0)}
\end{equation}
and
\[ B=\left\{\begin{array}{rcl}
           \coth{P/2}& \mbox{if}& \epsilon \epsilon' =1\\
           \tanh{P/2}& \mbox{if}& \epsilon \epsilon' =-1
           \end{array}\right. \] 
\end{itemize}

In all these expressions the soliton velocity was taken
 equal with unity and the solutions are written in Hirota's formalism.

The aim of this paper is to investigate a larger class 
of similar solutions. In the second paragraph starting from a certain Backlund 
transformation and its solutions found by Hirota and Satsuma 
\cite{hirsat},
a class of complex and implicitly nonsingular solutions is determined.
In the next one a general form of 1-rational N-soliton solution
is written down in a wronskian form
\cite{nimmo}. In the last section starting from 
a nonlinear superposition formula \cite{toda1}
\cite{wadati} the class of rational solutions
is enlarged, some of them being different from those found using AS limiting
procedure. It is observed that they satisfy also the Boussinesq equation
and it is conjectured that there is a larger class of common
rational solutions of both equations. This fact was somehow foreseen by
Gibbon \cite{gibbon}
 in his study of the poles of Toda lattice and the connection
with other Hamiltonian N-body systems and continuum limits as KdV and
Boussinesq equations.

\section{Rational Solutions}
The adimensional form of Toda equation \cite{hirsat}
 is given by:
\begin{equation}
\ddot y_{n}(t)=\exp{[-(y_{n}-y_{n-1})]}- \exp{[-(y_{n+1}-y_{n})]}
\label{todaheq}
\end{equation}
where $y_{n}(t)$ is the displacement of the n-th
particle from its equilibrium position.If we set:
\begin{equation}
\exp{[-(y_{n}-y_{n-1})]}=\frac{d^{2}}{d t^{2}} \ln{f_{n}(t)}
\end{equation}
and introducing in (\ref{todaheq})  Hirota form for Toda equation is obtained:
\begin{equation}
\left[D_{t}^{2}
-4\sinh^{2}{\left(\frac{D_{n}}{2}\right)}\right] f_{n}(t).f_{n}(t)=0
\label{todaeq}
\end{equation}			
where $D_{t}$ and $D_{n}$ are the well known Hirota bilinear operators.

Hirota and Satsuma \cite{hirsat}
 have introduced the following Backlund transformations
(BT) for Toda lattice and equivalent systems:
\begin{equation}
D_{t} f_{n}'.f_{n}+2\alpha \sinh{\left(\frac{D_{n}}{2}\right)} g_{n}'.g_{n}
=0 \nonumber
\end{equation}
\begin{equation}
D_{t} g_{n}'.g_{n}+2\alpha^{-1}
\sinh{\left(\frac{D_{n}}{2}\right)}f_{n}'.f_{n}=0\label{bt}
\end{equation}
\begin{equation}
\left[\beta_{1} \sinh{\left(\frac{D_{n}}{2}\right)}
+\cosh{\left(\frac{D_{n}}{2}\right)}\right]g_{n}'.g_{n}=0
\end{equation}
\begin{equation}
\left[\beta_{2} \sinh{\left(\frac{D_{n}}{2}\right)}
+\cosh{\left(\frac{D_{n}}{2}\right)}\right]f_{n}'.f_{n}=0
\end{equation}
where $\alpha$, $\beta_{1}$ and $\beta_{2}$ are constants
satisfying the relation:
				$$\alpha^{-1} (\beta_{1}^{2}-1)=\alpha(\beta_{2}^{2}-1)$$
 $(f_{n},f_{n}')$ and $(g_{n},g_{n}')$
are pairs of solutions of Toda equation of the same type.

The $1$-soliton and $2$-soliton solutions obtained from these BT are given by:

$$f_{n}=1+\exp{\left(2 \eta+\phi \right)}$$
\begin{equation}
f_{n}'=1+\exp{\left(2 \eta+\phi' \right)}\label{1sol}
\end{equation}
$$g_{n}=1+\exp{\left(2 \eta+\psi \right)}$$
$$g_{n}'=1+\exp{\left(2 \eta+\psi' \right)}$$

and
\begin{eqnarray}
  f_{n}=1+\exp{\left(2 \eta_{1}\!+\!\phi_{1}\right)}+
    \exp{\left(2 \eta_{2}\!+\!\phi_{2}\right)}+\exp{\left(2 (\eta_{1}
    \!+\!\eta_{2})\!+\!\phi_{1}\!+\!\phi_{2}\!+\!A_{1 2}\right)}
      \nonumber\\
  f_{n}'=1+\exp{\left(2 \eta_{1}\!+\!\phi_{1}'\right)}
    +\exp{\left(2 \eta_{2}\!+\!\phi_{2}'\right)}+
    \exp{\left(2(\eta_{1}\!+\!\eta_{2})
    \!+\!\phi_{1}'\!+\!\phi_{2}'\!+\!A_{1 2}\right)}
      \nonumber\\
  g_{n}=1+\exp{\left(2 \eta_{1}\!+\!\psi_{1}\right)}
    +\exp{\left(2 \eta_{2}\!+\!\psi_{2}\right)}
    +\exp{\left(2(\eta_{1}\!+\!\eta_{2})\!+\!\psi_{1}\!+\!\psi_{2}\!+
      \!A_{1 2}\right)}
      \nonumber\\
  g_{n}'=1+\exp{\left(2 \eta_{1}\!+\!\psi_{1}'\right)}
    +\exp{\left(2 \eta_{2}\!+\!\psi_{2}'\right)}
    +\exp{\left(2(\eta_{1}\!+\!\eta_{2})\!+\!\psi_{1}'\!+\!\psi_{2}'\!
    +\!A_{1 2}\right)}
\label{2sol}
\end{eqnarray}
Here $\epsilon_{i}=\pm 1$ depending on the
 propagation direction of the soliton.
$$\eta_{i}=\Omega_{i}t-P_{i}n+\eta_{i}^{(0)}$$
$$\Omega_{i}=\epsilon_{i}\sinh{P_{i}}$$
$$\exp{A_{1 2}}=\left[\frac{\epsilon_{1}\exp{P_{1}}-\epsilon_{2}\exp{P_{2}}}
{1-\epsilon_{1}\epsilon_{2}\exp{(P_{1}+P_{2})}}\right]^{2}$$
and
$$\exp{\phi_{i}}=\epsilon_{i}\alpha^{-1}\beta_{1}+\beta_{2}\cosh{P_{i}}-
                \sinh{P_{i}}$$
$$\exp{\phi_{i}'}=\epsilon_{i}\alpha^{-1}\beta_{1}+\beta_{2}\cosh{P_{i}}+
                 \sinh{P_{i}}$$
$$\exp{\psi_{i}}=\epsilon_{i}\alpha^{-1}(\epsilon_{i}\alpha\beta_{2}
                +\beta_{1}\cosh{P_{i}}-\sinh{P_{i}})$$
$$\exp{\psi_{i}'}=\epsilon_{i}\alpha^{-1}(\epsilon_{i}\alpha\beta_{2}
                +\beta_{1}\cosh{P_{i}}+\sinh{P_{i}})$$
A class of complex solutions are obtained with a very simple chioce
of$\alpha$, $\beta_{1}$ and $\beta_{2}$ namely:
$$\alpha=i$$
$$\beta_{1}=-\beta_{2}=1$$
In the limiting procedure of Ablowitz and Satsuma\cite{as1}
 the N-soliton solution is expanded in power series of $P$,
 $P_{i}=\delta P_{i}'$, with $\delta \rightarrow 0$
and the phase shifts $\exp{2\eta_{i}^{(0)}},i=1,...N$
 are determined cancelling all the terms of
order$O(P^{k})$ with $k$ at least smaller than $N$. Then the 
N-rational solution is found from the first nonvanishing term of order
$O(P^{k})$ with $k\geq N$.
\footnote{We have to take into account also that
$f$ and $f \exp{(\alpha t+\beta)}$ are two equivalent solutions with 
$\alpha$ , $ \beta$ does not depending on  $t$}

In this way starting from the 1-soliton solutions (\ref{1sol})
 and taking $\exp{2 \eta^{0}}={(1 \pm i)^{-1}}$
the 1-rational solutions are found from $O(P)$ terms
and are given by:
$$f_{n}=n-(1-i\epsilon)/4-\epsilon t$$
\begin{equation}
f_{n}'=n+(1-i\epsilon)/4-\epsilon t \label{1rats}
\end{equation}				
$$g_{n}=n+(1+i\epsilon)/4-\epsilon t$$
$$g_{n}'=n-(1+i\epsilon)/4-\epsilon t$$
They satisfy the following symmetry relations:
$$f_{n}=g_{n-1/2}, f_{n}'=g_{n+1/2}'$$
\begin{equation}
f_{n}=g_{n}'^{*}, f_{n}'=g_{n}^{*} \label{symmetry}
\end{equation}
where (*) means the complex conjugation. It is easily seen that they verify 
the BT (\ref{bt}) and the Toda equation (\ref{todaeq}). In fact any expression
				$$f_{n}=n \pm \epsilon t +z$$
with $z$ a complex number is a solution of (\ref{todaeq}).

In the case of 2-rational solutions two different expressions are obtained
depending if the two solitons are moving in opposite direction
 $(\epsilon_{1} \epsilon_{2}=-1)$
or in the same direction$(\epsilon_{1} \epsilon_{2}=1)$.
Thus if $\epsilon_{1} \epsilon_{2}=-1$ taking the limit $P_{1}\rightarrow 0
, P_{2}\rightarrow 0$
and choosing:
				$$\exp{2\eta_{1}^{(0)}}=(1+i)^{-1},
\exp{2\eta_{2}^{(0)}}=(1-i)^{-1}$$
the $O(1)$ and $O(P)$ terms cancel and the following
2-rational solutions are found from $O(P^{2})$ terms:
$$f_{n}=(n-1/4)^{2}-(t-i/4)^{2}-1/4$$
\begin{equation}
f_{n}'=(n+1/4)^{2}-(t+i/4)^{2}-1/4 \label{2rsolopposite}
\end{equation}
$$g_{n}=(n+1/4)^{2}-(t-i/4)^{2}-1/4$$
$$g_{n}'=(n-1/4)^{2}-(t+i/4)^{2}-1/4$$
They satisfy the symmetry relations (\ref{symmetry})
 and verify the BT (\ref{bt}) and Toda equation (\ref{todaeq}) .

When the two solitons are moving in the same direction
$\epsilon_{1} \epsilon_{2}=1$ the rational
solutions are obtained from the $O(P^{3})$
terms. As in our previous calculations \cite{carstea}
 the phase shifts have to be $P$
dependent and with the choice:
				$$\exp{2\eta_{1}^{(0)}}=
\frac{-1}{1+i\epsilon} \frac{P_{1}+P_{2}}{P_{1}-P_{2}}
-\frac{i\epsilon}{4}P_{1}P_{2}$$
$$\exp{2\eta_{2}^{(0)}}=\frac{1}{1+i\epsilon} \frac{P_{1}+P_{2}}{P_{1}-P_{2}}-
 \frac{i\epsilon}{4}P_{1}P_{2}$$
the terms of order $O(1)$, $O(P)$
 and$O(P^{2})$ are vanishing and from the $O(P^{3})$
ones we get:
$$f_{n}=4\left[n-(1-i\epsilon)/4-\epsilon t\right]^{3}
-\left[n-(1-i\epsilon)/4-\epsilon t\right]+2\epsilon t+i\epsilon/2$$
$$
f_{n}'=4\left[n+(1-i\epsilon)/4-\epsilon t\right]^{3}
-\left[n+(1-i\epsilon)/4-\epsilon t\right]+2\epsilon t-i\epsilon/2
$$
$$g_{n}=4\left[n+(1+i\epsilon)/4-\epsilon t\right]^{3}
-\left[n+(1+i\epsilon)/4-\epsilon t\right]+2\epsilon t+i\epsilon/2$$
$$g_{n}'=4\left[n-(1+i\epsilon)/4-\epsilon t\right]^{3}
-\left[n-(1+i\epsilon)/4-\epsilon t\right]+2\epsilon t-i\epsilon/2$$
Again the symmetry relations (\ref{symmetry})are satisfied and by 
straightforward 
calculations one checks that they are solutions of BT and Toda equations.

Higher order rational solutions can be found in the same way but the 
calculations become more and more difficult.

Beside the nice symmetry properties and their explicit form these solutions
are no more singular, although their physical significance is not yet clear.
  
\section{Mixed rational-soliton solution}

We shall study now the class of mixed rational- soliton solutions
also known as quasisolitons. As mentioned in the Introduction they were 
investigated firstly for KdV \cite{ac}, \cite{aair} 
 using different methods. Previously we have
shown that they exist also for Toda lattice \cite{carstea}.
The method used
 was the same "long wave limit".In the present section we shall extend these 
calculations and we shall present an explicit form for the
 1-rational N-soliton solution.
The starting point is the expression of the $\left(N+1\right)$ soliton 
solution written as a wronskian determinant. Following Nimmo's \cite{nimmo}
 notations
we have:
\begin{displaymath}
f_{n}^{(N+1)}(t)=\left|\begin{array}{cccc}
                       E_{1}^{+}+E_{1}^{-}&\frac {d}{dt} (E_{1}^{+}+E_{1}^{-})
                       &\ldots
                       &\frac{d^{N}}{dt^{N}} (E_{1}^{+}+E_{1}^{-})\\
                       E_{2}^{+}+E_{2}^{-}&\frac{d}{dt} (E_{2}^{+}+E_{2}^{-})
                       &\ldots
                       &\frac{d^{N}}{dt^{N}} (E_{2}^{+}+E_{2}^{-})\\
                       \vdots&\vdots&\ldots&\vdots\\
                       E_{N+1}^{+}+E_{N+1}^{-}
                       &\frac{d}{dt} (E_{N+1}^{+}+E_{N+1}^{-})&\ldots
                       &\frac{d^{N}}{dt^{N}} (E_{N+1}^{+}+E_{N+1}^{-})
                  \end{array}\right|
\end{displaymath}
where
				$$E_{i}^{\pm}=
a_{i}^{\pm}\exp{\left[\pm P_{i} n+\epsilon t \exp{(\pm P_{i})}\right]}
,  i=1,...N+1$$
Here $a_{i}^{\pm}$ are functions of the initial phase $\eta_{i}^{0}$ and
 $\epsilon_{i}= \pm 1$ 
depending on the propagation direction of the i-th soliton.
The 1-rational N-soliton is obtained in the limit
 $P_{N+1}\rightarrow 0$.This affects only
the last line of the determinant.Dropping the index $N+1$ we have to expand 
in power series of $P$ and keeping only the linear terms we get:
$$\frac{d^{j}}{dt^{j}} (E_{N+1}^{+}+E_{N+1}^{-})\rightarrow$$

$$\rightarrow \exp{(\epsilon t)}\epsilon^{j}\left[(a^{+}+a^{-})+
P(n+j+\epsilon t)(a^{+}-a^{-})
+O(P^{2})\right]$$
The vanishing of $O(1)$ terms gives:
				$$a^{+}+a^{-}=0$$
and the 1-rational N-soliton solution is obtained from the $O(P)$
terms, namely:
\begin{displaymath}
F^{1,N}(t)=\left| \begin{array}{ccccc}
E_{1}^{+}+E_{1}^{-}&\ldots&
\frac{d^{j}}{dt^{j}} (E_{1}^{+}+E_{1}^{-})&\ldots&\frac{d^{N}}{dt^{N}}
 (E_{1}^{+}+E_{1}^{-})\\
\vdots&\ldots&\vdots&\ldots&\vdots\\
E_{N}^{+}+E_{N}^{-}&\ldots&
\frac{d^{j}}{dt^{j}} (E_{N}^{+}+E_{N}^{-})&\ldots&\frac{d^{N}}{dt^{N}}
(E_{N}^{+}+E_{N}^{-})\\
n+\epsilon t&\ldots&\epsilon^{j}(n+j+\epsilon t)&
\ldots&\epsilon^{N}(n+N+\epsilon t)
\end{array}\right|
\end{displaymath}
Higher order rational-N soliton solutions can be found starting from Wronskian
determinants  of order $(N+2)$, $(N+3)$ and 
taking the limit $P_{N+1}\rightarrow 0, P_{N+2}\rightarrow 0$

\section{Extended class of rational solutions}

One of the main advantages of using the method of Backlund transformations
is the possibility to derive a nonlinear superposition formula \cite{wadati}.
For Toda
lattice starting from a given solution $y_{n}^{(0)}$ through
two distinct BT the solutions $y_{n}^{(1)}$ and 
$y_{n}^{(2)}$ can be constructed.
Interchanging the BT and assuming the permutability of the Bianchi
diagram a new solution $y_{n}^{(1,2)}$ is expressed in terms of 
$y_{n}^{\left(0\right)}$,$y_{n}^{\left(1\right)}$, $y_{n}^{\left(2\right)}$
by a simple algebraic relation \cite{wadati}:
\begin{eqnarray}
&~&{\exp{(y_{n}^{(1,2)}-\gamma^{(1,2)}+y_{n}^{(0)}-\gamma^{(0)})}=}
\nonumber\\
&~& =\exp{(y_{n+1}^{(1)}-\gamma^{(1)}+y_{n+1}^{(2)}-\gamma^{(2)})}
 \frac{z_{1}\exp{(y_{n}^{(2)}-\gamma^{(2)})}
-z_{2}\exp{(y_{n}^{(1)}-\gamma^{(1)})}}
{z_{1}\exp{(y_{n+1}^{(2)}-\gamma^{(2)})}-z_{2}\exp{(y_{n+1}^{(1)}-
\gamma^{(1)})}}
\nonumber\\
&~&\label{superposition formula}
\end{eqnarray}
Here
$$\gamma^{(0)}=y_{-\infty}^{(0)},           \gamma^{(1)}=y_{-\infty}^{(1)}$$
$$\gamma^{(2)}=y_{-\infty}^{(2)},          \gamma^{(1,2)}=y_{-\infty}^{(1,2)}$$
and
$$z_{1}=A_{1}\exp{(\gamma^{(0)}-\gamma^{(1)})}=A_{1}\exp{(\gamma^{(2)}-
\gamma^{(1,2)})}$$
$$z_{2}=A_{2}\exp{(\gamma^{(0)}-\gamma^{(2)})}=A_{2}\exp{(\gamma^{(1)}-
\gamma^{(1,2)})}$$
where $A_{1}$, $A_{2}$ are constants.\\
Then we have
$$\exp{(y_{-\infty}^{(i)}-y_{n}^{(i)})}=\prod_{k \leq n}^{-\infty} 
\exp{(y_{k-1}^{(i)}-y_{k}^{(i)})}=\frac{f_{n+1}^{(i)}}{f_{n}^{(i)}}
\lim_{k\rightarrow -\infty}\frac{f_{n-k}^{(i)}}{f_{n-k+1}^{(i)}}$$
It is convenient to consider:
$$\lim_{k\rightarrow -\infty}
\frac{f_{n-k}^{(i)}}{f_{n-k+1}^{(i)}}=1$$
Then the superposition formula (\ref{superposition formula}) becomes:
\begin{equation}
\frac{f_{n}^{(1,2)}}{f_{n+1}^{(1,2)}}=\frac{z_{1}f_{n}^{(2)}f_{n+1}^{(1)}-
z_{2}f_{n}^{(1)}f_{n+1}^{(2)}}{z_{1}f_{n+1}^{(2)}f_{n+2}^{(1)}-
z_{2}f_{n+1}^{(2)}f_{n+2}^{(1)}}  \frac{f_{n+2}^{(0)}}{f_{n+1}^{(0)}}
\label{nsf}
\end{equation}
It can be used not only to generate higher order soliton solutions 
but also other types of solutions. We shall apply it to find a larger class of 
2-rational solutions starting from $f_{n}^{(0)}=const.$. It is convenient
and possible to use a simplified form of (\ref{nsf}) namely
\begin{equation}
f_{n}^{(1,2)}=(z_{1}f_{n}^{(2)}f_{n+1}^{(1)}-z_{2}f_{n}^{(1)}f_{n+1}^{(2)})/
f_{n+1}^{(0)} \label{simplified nsf}
\end{equation}
Taking $f_{n}^{(1)}$ and $f_{n}^{(2)}$
  as the 1-rational solution			
$$f_{n}^{(1)}=n+t+\lambda_{1},        f_{n}^{(2)}=n-t+\lambda_{2}$$
and considering $z_{1}+z_{2}=0$ one obtains:
\begin{equation}
f_{n}^{(1,2)}=n^{2}-t^{2}+(\lambda_{1}+\lambda_{2}+1)n+(\lambda_{1}-\lambda_{2}
)t+(\lambda_{1}\lambda_{2}+\frac{\lambda_{1}+\lambda_{2}}{2})  \label{gen2rat}
\end{equation}
It is easily shown that for any complex numbers $\lambda_{1},\lambda_{2}$
  , $f_{n}^{(1,2)}$ is a solution of Toda lattice and with an appropriate 
choice of $\lambda_{1}$ and $\lambda_{2}$ namely $\lambda_{1}=\lambda_{2}=-1/2$
the already known 2-rational solution(\ref{zakrs2-})
 obtained by the "long wave limit" method of Ablowitz and Satsuma is found.

In fact any polinomial form $n^{2}-t^{2}+an+bt+c$ such that $a^{2}-b^{2}=
4c+1$  is a solution for Toda lattice where $a$, $b$ and $c$ are complex 
numbers.
The solution(\ref{gen2rat}) verifies identically these conditions.One can
see that for 
$$\lambda_{1},\lambda_{2} \in \{(i-3)/4,
 -(i-3)/4, (i-1)/4, -(i-1)/4 \}$$ 
we can recover the set of complex 2 rational
 solutions(\ref{2rsolopposite}). The above superposition formula is not useful
 to find also 
rational solutions written as 3-degree polinomials in $n$ and $t$ starting
from 1-rational solutions. Perhaps more complicated higher order rational 
solutions can be found using it in the form of ratios of polinomials, but 
soon the calculations become very tedious.

On the other hand it is worthily to mention the similarity of these solutions
with the coresponding rational solutions of Boussinesq equation:
$$(D_{t}^{2}-D_{x}^{2}-1/12D_{x}^{4})f(x,t).f(x,t)=0$$.
For this variant of Boussinesq equation the rational solutions	are :
$$x \pm t$$
$$x^{2}-t^{2}-1/4$$
$$4(x \pm t)^{3}-(x \pm t) \mp 2t$$
They have been found by Ablowitz and Satsuma using the "long wave limit" method
\cite{as1}.Also, we can see that:
$$x \pm t+\lambda$$
$$x^{2}-t^{2}+ax+bt+c$$ with $$ a^{2}-b^{2}=4c+1$$
are solutions for Boussinesq equation. As this equation is one of the continuum
limit of the Toda lattice we can ask ourselves if there are more identical 
rational solutions of both equations. We conjecture that this happens, although
they are not neccesary polinomials in Hirota formalism.

\end{document}